\documentclass[a4paper, 12pt]{article}

\usepackage[margin=1in,includefoot]{geometry}
\usepackage[english]{babel}	
\usepackage[utf8]{inputenc}	
\usepackage[T1]{fontenc} 
\usepackage{textcomp} 
\usepackage{setspace} 
\usepackage{enumitem} 
\usepackage{amsmath}
\usepackage{accents} 
\usepackage[usenames, dvipsnames]{xcolor}
\usepackage{url}

\usepackage[colorlinks=true, linkcolor=blue, urlcolor=black, bookmarksopen=true]{hyperref}
\usepackage{bookmark} 
\usepackage[dvipsnames]{xcolor}	

\usepackage{graphicx}	
\usepackage{subfig} 
\usepackage{float}  
\usepackage{bbm}
\usepackage[round]{natbib}
\usepackage{booktabs}		
\usepackage{authblk}
\usepackage[textfont=it, labelfont=bf, font=small]{caption}
\usepackage{comment}

\title{\bfseries New approach for stochastic downscaling and bias correction of daily mean
temperatures to a high-resolution grid} 
\author[1,2]{Qifen Yuan}
\author[3]{Thordis L. Thorarinsdottir}
\author[1]{Stein Beldring}
\author[1]{Wai Kwok Wong}
\author[1]{Shaochun Huang}
\author[2]{Chong-Yu Xu}

\affil[1]{Norwegian Water Resources and Energy Directorate, Oslo, Norway}
\affil[2]{University of Oslo, Oslo, Norway}
\affil[3]{Norwegian Computing Center, Oslo, Norway} 

\date{\today}

\begin{document}
\maketitle

%%%%%%%%
% Abstract
%%%%%%%%
\begin{abstract}
	\noindent
In applications of climate information, coarse-resolution climate projections commonly need to be downscaled to a finer grid. One challenge of this requirement is the modeling of sub-grid variability and the spatial and temporal dependence at the finer scale. Here, a post-processing procedure is proposed for temperature projections that addresses this challenge. The procedure employs statistical bias correction and stochastic downscaling in two steps. In a first step, errors that are related to spatial and temporal features of the first two moments of the temperature distribution at model scale are identified and corrected. Secondly, residual space-time dependence at the finer scale is analyzed using a statistical model, from which realizations are generated and then combined with appropriate climate change signal to form the downscaled projection fields. Using a high-resolution observational gridded data product, the proposed approach is applied in a case study where projections of two regional climate models from the EURO-CORDEX ensemble are bias-corrected and downscaled to a $1 \times 1$ km grid in the Tr{\o}ndelag area of Norway. A cross-validation study shows that the proposed procedure generates results that better reflect the marginal distributional properties of the data product and have better consistency in space and time than empirical quantile mapping.

\vspace{3mm}
\noindent
{\em Keywords}: Climate model output, local variability, model output statistics, post-processing, space-time consistency, weather generator.    
\end{abstract}

%%%%%%%%%%
%  Introduction
%%%%%%%%%%

\section{Introduction}\label{sec:introduction}

Climate change impacts often realize at local to regional scales, resulting in impact models such as hydrological models, forest growth models and crop models requiring tailored information on future climate at fine spatial and temporal scales \citep{barros2014ipcc, Hanssen-Bauer&2017}. In particular, many of these impact models are conducted on a very fine spatial grid and at daily timescale \citep[e.g.][]{Beldring&2003}. Future climate information commonly derives from coupled atmosphere-ocean general circulation models (GCMs) that currently neither provide unbiased nor local to regional scale information. Regional climate models (RCMs), with a spatial resolution of 10-15 km, provide a partial bridge for the spatial scale gap. While ensembles of RCMs are able to capture basic features of regional climate variability in space and time \citep{Kotlarski&2014}, their output may still contain substantial errors, partly inherited from the driving GCM \citep{Rummukainen2010, Hall2014}. 

Impact studies are generally performed by comparing results for a reference climate to those obtained under a projected future climate. Where high-resolution gridded climate input data are required, the reference results are commonly based on gridded data products derived from lower dimensional observations such as a network of surface observation stations \citep[e.g.][]{Lussana&2018a, Lussana&2018b}. These data products come with their own inherent biases which are difficult to correct due to a lack of data. For an accurate assessment of climate impact, one goal is thus to generate high-resolution realizations of future climate which properties differ from those describing the reference climate only in terms of the expected climate change between the two time periods. In particular, statistical aspects such as the space-time variability and dependence at the finer scale should be realistically represented \citep{Wood&2004, Beldring&2008}. 

To generate tailored climate information for various impact studies, post-processing methods are almost routinely performed on climate model outputs. In a recent monograph on the subject, \citet{MaraunWidmann2018} identify three classes of statistical post-processing methods. {\em Model output statistics} (MOS) approaches apply a statistical transfer function between simulated and observed data, and are employed for both bias correction and downscaling. Depending on the specific needs of the climate information user, a wide variety of such methods are in use, ranging from simple mean adjustment to flexible, potentially multivariate quantile mapping methods \citep{Maraun&2010, Piani2012, Vrac&2012, VracFriederichs2015, Cannon2016, Vrac2018}. For downscaling, {\em perfect prognosis} (PP) methods establish a statistical link between large-scale predictors and local-scale predictands typically in a regression framework, while {\em weather generators} (WGs) are stochastic models that explicitly model marginal and higher order structures.  WGs are widely used for generating weather time series at stations \citep{Semenov&1997}, with some extensions to multi-site \citep{Wilks1999, Wilks2009} and multivariate \citep{Kilsby&2007} settings.

One common issue with MOS methods applied to downscaling is that they are not able to capture spatial and temporal variability at the finer scale \citep[e.g.][]{Maraun&2017, MaraunWidmann2018}. The transfer functions derived from the historical period are transformations of the stochasticity at the model scale which are often not realistic at the required fine scale. Hence, including stochastic components into the bias correction procedure is imperative to account for local-scale variability \citep{Maraun&2017}. Recently proposed stochastic downscaling methods have proven skillful in modeling the small-scale variability of precipitation occurrence and intensity across sets of point locations \citep{Wong&2014, Volosciuk&2017}. The impact models considered in our application commonly require high-resolution gridded input and thus approaches that scale to high dimensional and spatially coherent settings.

Further, it has been argued that methods based on PP assumptions where it is assumed that daily based coarse-scale information can be used to predict the probability distribution at the local-scale are not appropriate for free-running model simulations such as the RCMs from CORDEX \citep{Jacob&2014}. For full-field downscaling without PP assumptions, techniques of shuffling the time series produced by univariate bias correction have been proposed \citep[e.g. ``Schaake shuffle''][]{Clark&2004}, both for temporal \citep{VracFriederichs2015}, and multi-site and multivariate reordering \citep{Vrac2018}. The shuffling techniques impose historical rank correlation structure on the bias-corrected data. They have, in some instances, been shown to underrepresent the dependence structure \citep{Vrac2018}. Moreover, the size of the shuffled data set is restricted to the size of the observational data set.

Alternatively, multi-site WGs that explicitly model the fine-scale stochasticity are able to generate spatially and temporally coherent fields and thus have shown potential for full-field downscaling \citep{Wilks2010, Wilks2012}. This approach, however, has been typically applied where parameters are calibrated first at single locations and then interpolated onto a grid consisting of a small set of grid points \citep{Wilks2009};  which is not straightforward to work with when gridded data products are available and have been used to train the impact models. Besides, they are primarily constructed for generating daily precipitation, whereas daily mean temperature has its own properties \citep{Huybers&2014} and is an equally important input to e.g. hydrological models \citep{Xu1999}. Our objective is thus to propose a full-field downscaling approach for daily mean temperature that explicitly accounts for the fine-scale variability and dependence in both space and time. 

Specifically, we introduce a two-stage statistical post-processing procedure that bias-corrects and downscales RCM simulations to a high-resolution grid. In the first stage, a MOS approach is applied to bias-correct RCM output at the model scale by comparing it against upscaled gridded data product. Daily mean temperatures are generally considered well represented by a Gaussian distribution \citep[e.g.][]{Piani&2010}. Here, we apply a transfer function where the parameters of the Gaussian distribution vary across space and time to account for seasonal and geographic changes in temperatures. Secondly, we construct a WG to simulate pseudo-observations that replicate the properties of a fine-scale gridded data product under a stationary climate. Using a separable space-time correlation structure, the method is able to efficiently generate high dimensional realizations. We then impose these realizations with appropriate climate change signal derived from the RCM simulations. The approach can thus be thought of as a delta change method that preserves space-time consistency. 

The remainder of the paper is organized as follows. In Section~\ref{sec:data} we describe the data and the study area. In the next Section~\ref{sec:methodology} we describe the proposed post-processing procedure and briefly discuss evaluation methods. Results are presented in Section~\ref{sec:results}, and the final Section~\ref{sec:conclusions} provides discussion and conclusions.

%%%%%%%%%%%%%%%
% Data and study area
%%%%%%%%%%%%%%%
\section{Data and study area}\label{sec:data} 

We apply our methodology to daily mean temperature simulations from two RCMs from the EURO-CORDEX-11 ensemble. One combines the COSMO Climate Limited area Model (CCLM) from the Potsdam Institute for Climate Research \citep{Rockel&2008} with boundary conditions from the CNRM-CM5 Earth system model (referred to as RCM1 in the following text) developed by the French National Centre for Meteorological Research \citep{Voldoire&2013}, whereas the other (referred to as RCM2) combines the CCLM model with boundary conditions from the MPI Earth system model developed by the Max Planck Institute for Meteorology \citep{Giorgetta&2013}. The RCM simulations are conducted over the European domain at a spatial resolution of 0.11 degrees or about 12.5 km grid resolution \citep{Jacob&2014}.  In the historical period up to 2005 the outputs are simulated based on recorded emissions and are thus comparable to observed climate. 

For observational reference data, we use the seNorge gridded data product version 2.1 produced by the Norwegian Meteorological Institute \citep{Lussana&2018b} and available at \url{http://www.senorge.no}. The data result from an optimal spatial interpolation method applied to measurements at around 600 weather and climate stations for the period 1957 to present and are available at a spatial resolution of 1 km over an area covering the mainland Norway and an adjacent strip along the Norwegian border. For bias-correcting the RCM output, we upscale the seNorge data to the RCM grid by taking a weighted average over all seNorge grid cells found within each RCM grid cell, where the weights are area ratios of the seNorge cells to that RCM cell. 

\begin{figure}[H]
  \begin{center}
    \includegraphics[width=0.85\textwidth]{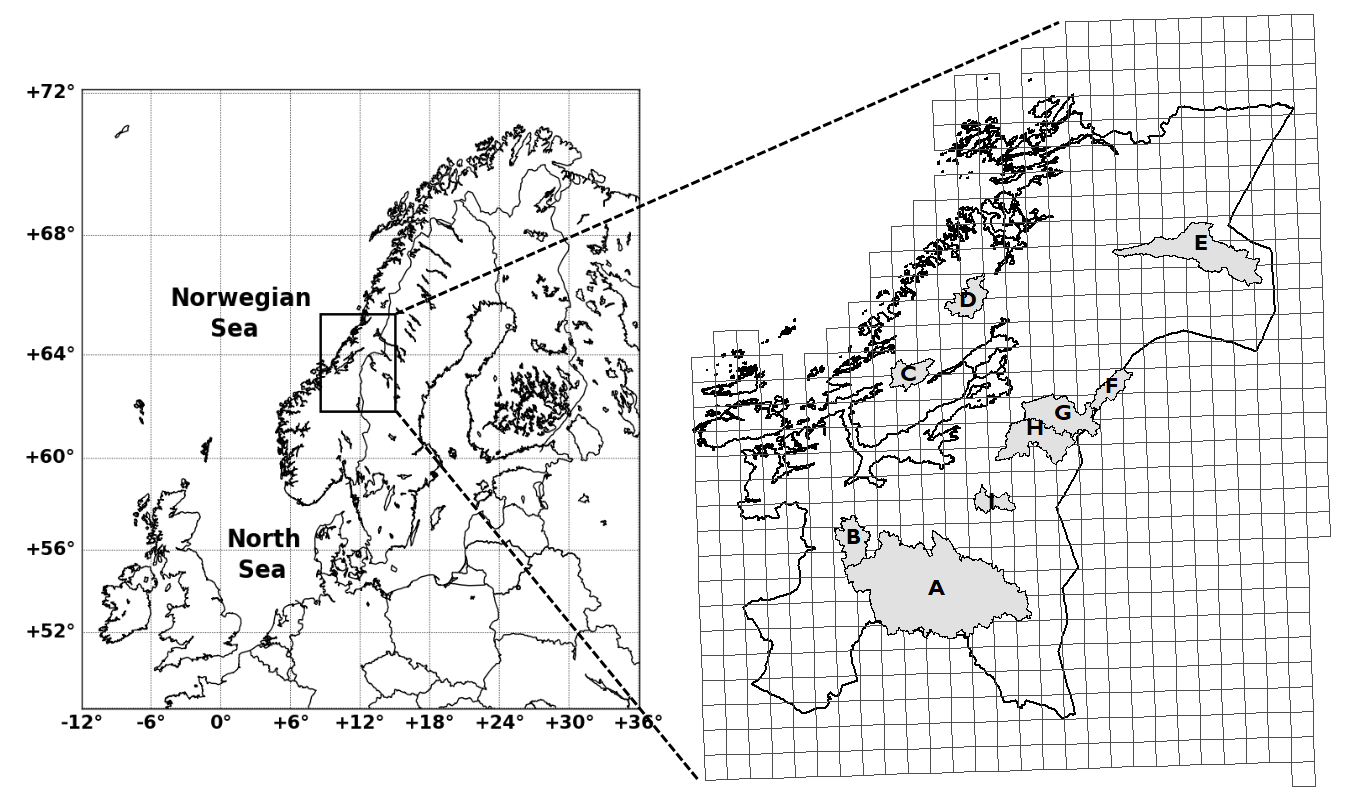}
    \caption{Our study area comprises the area of Tr{\o}ndelag in central Norway. For the RCM bias correction, we consider the entire Tr{\o}ndelag and a small part of neighboring Sweden, an area with 695 RCM grid cells (rectangular-like polygons) and 109 514 seNorge grid cells (within the polygons, not shown). For the stochastic downscaling, we consider nine hydrological catchments within Tr{\o}ndelag with catchment areas from 143 km$^2$ to 3086 km$^2$ (shaded in gray), see also Table~\ref{tab:catchments}.}
    \label{fig:studyarea}
  \end{center}
\end{figure}

For study area, we consider the Tr{\o}ndelag area in central Norway, see Figure~\ref{fig:studyarea}. The area comprises 695 RCM grid cells and 109 514 seNorge grid cells. The bias correction is performed over the entire study domain while the statistical downscaling focuses on nine hydrological catchments within the domain, see Figure~\ref{fig:studyarea} and Table~\ref{tab:catchments}. Two catchments Krinsvatn and Oeyungen have maritime climate while the rest have continental climate. For each catchment, the downscaling is performed over all seNorge grid cells within the RCM grid cells that cover the catchment. The spatial dimensions of the downscaling areas thus vary between approximately 940 and 5500 grid cells at 1 km resolution.  Both historical RCM simulations and seNorge observations are available over the time period 1957-2005. We use the time period 1957-1986 as a training period to estimate the parameters of the post-processing approaches and perform an out-of-sample evaluation over the remaining 19 years 1987-2005. As a result, the training period consists of 10 950 days while the test period comprises 6 935 days. 

\begin{table}[H]
  \caption{Characteristics of the nine hydrological catchments in Tr{\o}ndelag, Norway considered in the stochastic downscaling.}\label{tab:catchments}
  \centering
  \begin{tabular}{llrrr}
    \toprule
    Catchment & ID & Size 	& Downscaling area & Median elevation 	\\
    & & (km$^2$) & (km$^2$) & (m.a.s.l) \\
    \midrule
    Gaulfoss	 &A 	& 3086	& 5479	& 734 \\ 
    Aamot  	 &B		& 283  & 1112 & 460 \\ 
    Krinsvatn&C		& 206	& 1108 & 349 \\
    Oeyungen &D	& 239 	& 952   & 295 \\
    Trangen	 & E	& 852	& 2327 & 558 \\
    Veravatn & F	&175	& 1101 & 514 \\
    Dillfoss    &G	& 483	& 1863 & 506 \\
    Hoeggaas &H	& 495	& 1853 & 505 \\
    Kjeldstad  &I	& 143	& 940  & 578 \\
    \bottomrule
  \end{tabular}
\end{table}

Additionally, we use explanatory variables, or covariates, to describe the spatial variations in the statistical characteristics of the daily mean temperature distributions. We consider latitude, longitude and elevation as potential geographic covariates. Elevation information for the seNorge data is obtained from a digital elevation model based on a 100 m resolution terrain model from the Norwegian Mapping Authority \citep{Mohr2009}. We upscale these data in the same manner as the daily mean temperatures to obtain the elevation at the RCM scale. Note that this is not equal to the orography information provided by EURO-CORDEX.

%%%%%%%%%%%%%%
%%    Methodology                                            
%%%%%%%%%%%%%%

\section{Methods}\label{sec:methodology}

\begin{figure}[H]
	\begin{center}
		\includegraphics[width=0.75\textwidth]{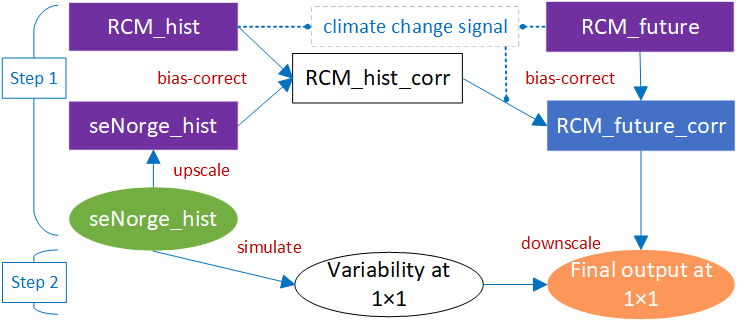}
		\caption{Proposed general framework for post-processing of climate model output.}
		\label{fig:modelFramework}
	\end{center}
\end{figure}

We propose a two-step post-processing approach for statistical bias correction and stochastic downscaling as demonstrated in Figure~\ref{fig:modelFramework}. First, biases at the RCM scale are identified and corrected, where the climate change signal simulated by RCM is preserved.  Then, we estimate the space-time residual variability at the finer seNorge scale using a statistical model, and, by simulating from this model, we are able to generate a set of realizations of a stationary future climate possessing the same space-time structures as the historical data product. Based on these, we compare three different approaches for adding layers of the climate change signal to obtain the final output. A detailed description of each step is given below.

\subsection{Bias correction}\label{subsec:biasCorrection}

For bias-correcting the RCM output, we perform weighted upscaling of the seNorge data product as described in Section~\ref{sec:data}. The RCM-simulated grid cell values represent areal averages; the upscaled seNorge data should thus be comparable to the RCM output in distribution. We follow e.g. \citet{Piani&2010} and assume that temperature can be modeled by a Gaussian distribution. However, rather than modeling each month separately, the parameters of the distribution are assumed to change smoothly across time and space.

Specifically, denote by $Y_{rt}$ the daily mean temperature in grid cell $r \in \{1, \ldots, R\}$ at time $t \in \{1, \ldots, T\}$, where $R$ denotes the number of grid cells and $T$ the number of days in a given RCM-scale data set. We then set  

\begin{equation}\label{eq:Y}		
  Y_{rt} \sim \textup{N} (\mu_{rt}, \sigma^2_{rt}),  
\end{equation}
where
\begin{align}
    \mu_{rt} & = f_1^{\mu}(\boldsymbol{c}_r) + f_2^{\mu}(t) + f_3^{\mu}(t), \label{eq:mu} \\
    \log(\sigma_{rt}) & = f_1^{\sigma}(\boldsymbol{c}_r) + f_2^{\sigma}(t), \label{eq:logSigma}
\end{align}
with
\begin{align}
    f_1^{\zeta}(\boldsymbol{c}_r) & = \alpha_{11} + \alpha_{12} c_{r1} + 
    \alpha_{13} c_{r2} + \alpha_{14} c_{r3}, \label{eq:f1} \\
    f_2^{\zeta}(t) & = \alpha_{21} \cos \left(\frac{2\pi \, d(t)}{365}\right) + 
    \alpha_{22} \sin \left(\frac{2\pi \, d(t)}{365}\right) \notag \\
	& \quad + \alpha_{23} \cos \left(\frac{4\pi \, d(t)}{365}\right) + 
	\alpha_{24} \sin \left(\frac{4\pi \, d(t)}{365}\right), \label{eq:f2} \\
    f_3^{\mu}(t) & = \alpha_{3} y(t), \label{eq:f3} 
\end{align}
for $\zeta \in \{ \mu, \sigma\}$. Here, $f_1$ models the spatially varying baseline of the two moments with $\boldsymbol{c}_r = (c_{r1}, c_{r2}, c_{r3})$ being latitude, longitude and mean elevation of grid cell $r$. Seasonal changes in the moments are captured by $f_2$ where $d(t)$ returns the calendar day of time point $t$ and $f_3$ describes potential linear trend in the first moment with $y(t)$ returning the calendar year normalized so that $\alpha_3$ describes the trend in degrees per decade.  

The model specified in equations (\ref{eq:Y})-(\ref{eq:f3}) has 17 coefficients, i.e. 9 coefficients for the first moment and 8 coefficients for the second moment. A two-step analysis of three RCM-scale data sets  where the mean and the residuals were analyzed separately found all the coefficients significant. For the bias correction we estimate all 17 coefficients for each data set simultaneously by numerically obtaining the maximum likelihood estimator (MLE) using the function \texttt{lmvar()} from the R \citep{R} package \texttt{lmvar} \citep{lmvar}. Subsequently, we adjust the estimated model parameters for the RCM simulations in the out-of-sample test period based on the estimates from the upscaled seNorge data and the RCM data in the training period, and the RCM-simulated changes from the training period to the test period. In particular, the correction (\textbf{Corr}) is similar to equation (12.7) in the book \citep{MaraunWidmann2018}, with two differences though: first, the mean and standard deviation terms are estimates by equations (\ref{eq:mu}) and (\ref{eq:logSigma}) varying across space and time; second, the standardized RCM anomalies (i.e. dividing by its own standard deviation) are rescaled by the square root of the variance of the upscaled seNorge data plus the RCM-simulated change in the variances between the two periods.  

For comparison, we consider two simple bias correction methods commonly used as benchmarks \citep[e.g.][]{RaisanenRaty2013} where only the mean of the RCM output is corrected using one common correction term across the entire domain (\textbf{Simple}) or independently for each grid cell (\textbf{LocalSimple}). These methods explicitly preserve the change in the long-term mean simulated by RCM.

\subsection{Stochastic downscaling}\label{subsec:statisticalDownscaling}
\subsubsection{Stationary space-time high-resolution model}\label{subsubsec:statisticalModel}

We model the space-time variability at the finer 1 km scale by a stochastic model that assumes stationarity and space-time separability in the residuals. In order to warrant these rather strict assumptions--assumed for computational feasibility--we estimate the model independently for each catchment, allowing for e.g. changes in the space-time variability across different climatic zones.

Let $X_{st}$ denote the daily mean temperature at time $t \in \{1, \ldots, T\}$ in the training period and fine-scale grid cell $s \in \{1, \ldots, S\}$ for a given catchment. We fit a model of the form given in (\ref{eq:Y})-(\ref{eq:f3}) above to this data set and generate the corresponding residuals
\begin{equation}\label{eq:Z}
  Z_{st} = \frac{X_{st} - \hat{\mu}_{st}}{\hat{\sigma}_{st}}.
\end{equation}
  We then estimate a residual model of the form 
\begin{align}
    Z_{st} & = \eta_t + \nu_{st}, \label{eq:Zmodel} \\
    \eta_t & \sim \textup{SN}(\mu_t, \sigma_{1t}, \sigma_{2t}), \label{eq:SN} \\
    U_t & = \Phi^{-1}(F_{\textup{SN}}(\eta_t)) \sim \textup{ARMA}(p, q), \label{eq:ARMA} \\
    \boldsymbol{\nu}_{t} & \sim \textup{N}(\boldsymbol{0}, \pmb{\mathsf{\Sigma}}_t), \label{eq:nu} \\     
    \textup{Cov}(\nu_{st}, \nu_{s't}) & = \theta_{0t} \mathbbm{1}\{\|s - s'\| = 0\} + \theta_{1t} \exp( - \| s - s'\|/\theta_{2t}). \label{eq:Cov}
\end{align}
Here, $\textup{SN}$ stands for the split normal distribution \citep{Wallis2014}, sometimes called the two-piece normal distribution, a three parameter generalization of the normal distribution that allows for asymmetry in the tails in that a separate scale parameter is used for each of the two tails.

\begin{figure}[H]
	\begin{center}
	\includegraphics[width=0.9\textwidth]{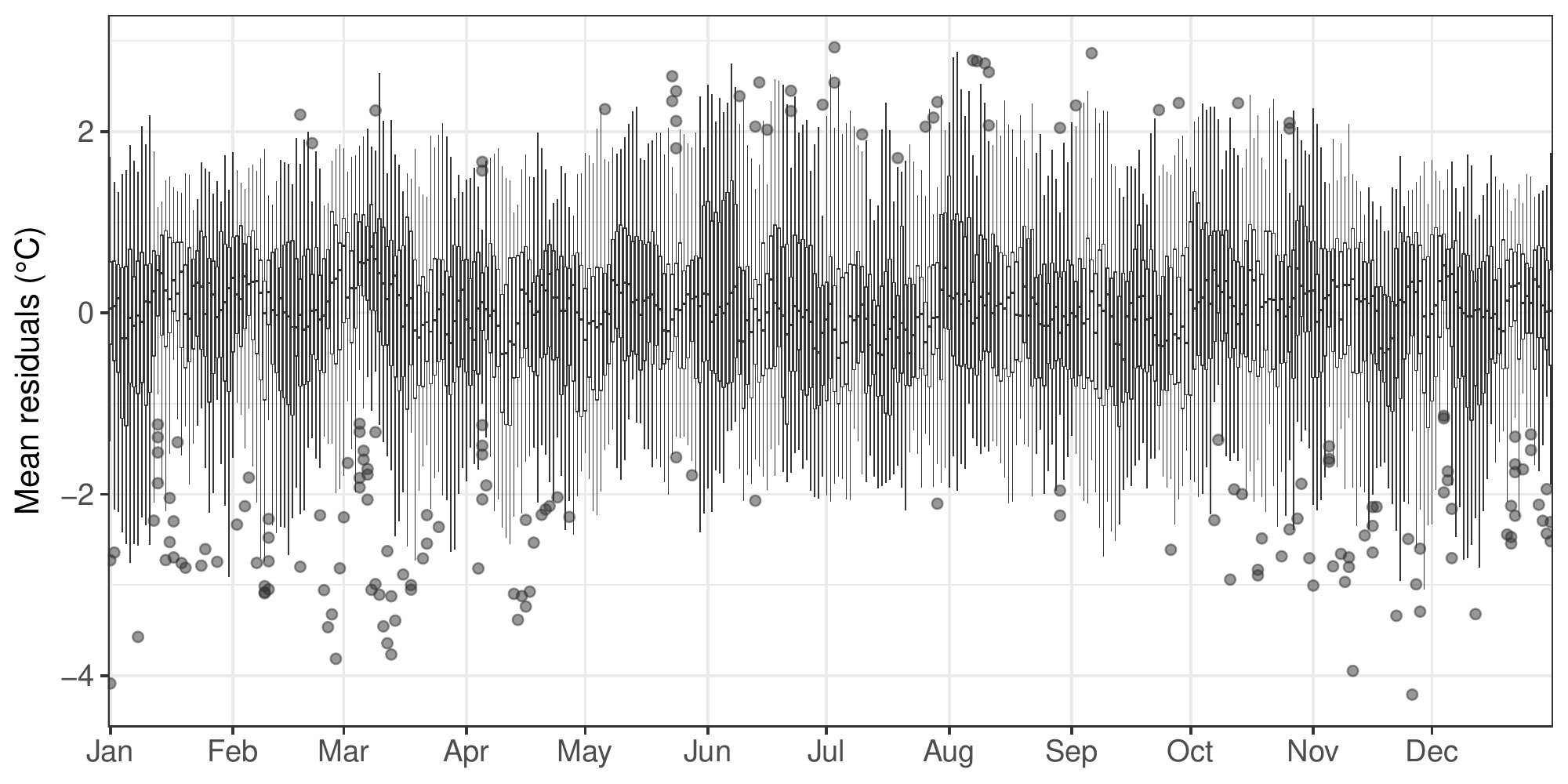}
	\caption{Boxplots of the mean residuals $\bar{Z}_{\cdot t} = \frac{1}{S} \sum_{s=1}^{S} Z_{st}$ where the mean is taken over all grid cells in catchment A which is the largest in our study area. The boxplot for each calendar day consists of all values for that calendar day in the training period 1957-1986.}
	\label{fig:boxplotEta}
	\end{center}
\end{figure}

We assume the residual field $\boldsymbol{Z}_t$ varies around a mean value of zero, so that the time series $\bar{Z}_{\cdot t} =  \frac{1}{S} \sum_{s=1}^{S} Z_{st}$ represents a reasonable approximation of the true temporal dependence. Although appearing stationary, the time series $\{ \bar{Z}_{\cdot t} \}$ does not seem to have Gaussian marginals as shown in Figure~\ref{fig:boxplotEta}. In particular, the marginals have a positive skewness in the warmer months and a negative skewness in the colder months. We thus employ a copula approach to estimate the temporal correlation \citep{Nelsen2007} where we combine split normal marginals (\ref{eq:SN}) and an autoregressive moving average (ARMA) structure (\ref{eq:ARMA}) to account for both the marginal skewness and the temporal correlation. Daily varying parameter estimates for the split normal distribution are obtained by numerically optimizing the likelihood function \texttt{logs\_2pnorm()} from the \texttt{R} package \texttt{scoringRules} \citep{Jordan&2018}. Subsequently, an ARMA model of order $(p,q)$ is estimated using the function \texttt{auto.arima()} from the \texttt{R} package \texttt{forecast} \citep{Hyndman&2018}. This approach optimally determines the values of $p$, $q$ and estimates the associated parameters. We found that $p=2, q=3$ fits the best for catchment A, $p=3, q=2$ for catchment F, and $p=q=2$ for the rest. For comparison, we have also investigated a simpler model with Gaussian marginals. However, this results in substantially reduced performance, see Section~\ref{subsubsec:resultsSDIQD}. 

Next, we assume that the spatial residuals $\boldsymbol{\nu}_t = \boldsymbol{Z}_t - \hat{\eta}_t$ follow a multivariate normal distribution \eqref{eq:nu} with a mean vector zero and a variance matrix specified by a stationary and isotropic covariance function of the exponential type \eqref{eq:Cov} so that the spatial correlation between two grid cells $s, s' \in \{1, \dots, S\}$ is determined by their Euclidean distance $\| s -s' \|$ \citep[e.g.][]{CressieWilke2015}. To estimate the parameters of the covariance function \eqref{eq:Cov}, we employ semi-variogram function given by
\begin{equation}\label{eq:vgmMod}
  \gamma_{\boldsymbol{\theta_t}} (h) = 
  \theta_{0t} + \theta_{1t} [1-\exp(-h/\theta_{2t}) ] 
\end{equation}
with $h = \| s - s' \|$. To account for potential seasonal changes in the spatial correlation structure, we obtain separate estimates for each month and, subsequently, fit a smooth function through each set of estimates to obtain smoothly changing daily estimates.

Denote by $\mathcal{T}$ the set of all time points from a given month with $|\mathcal{T}|$ the number of days in this set, and by $\mathcal{S}(h)$ the set of grid cell pairs that have distances within some small interval approximately centered around $h$ with $|\mathcal{S}(h)|$ the number of pairs in the set. We then estimate the covariance parameters by fitting the semi-variogram function \eqref{eq:vgmMod} to the empirical semi-variogram
\begin{equation} \label{eq:empVgm}
	\hat{\gamma}(h, \mathcal{T}) = \frac{1}{2|\mathcal{S}(h)|\, |\mathcal{T}|} 
	\sum \limits_{(s, s') \in \mathcal{S}(h)} 
	\sum \limits_{t \in \mathcal{T}} (\nu_{st}- \nu_{s't})^{2}. 
\end{equation}
Here, we employ the \texttt{R} package \texttt{spacetime} \citep{Pebesma2012} to organize our spatio-temporal residuals and \texttt{gstat} \citep{Pebesma2004, Graeler&2016} to calculate empirical semi-variograms and perform the fitting using decreasing weights on the pairs that are further apart. 

We can then simulate a set of residuals $Z_{st}^\ast$ for $s \in \{1,\ldots, S\}$ and $t \in \{1, \ldots, T\}$ in four steps:
\begin{enumerate}
\item Simulate	$U_t^\ast \sim \widehat{\textup{ARMA}}(p, q)$; 
\item Set $\eta_t^\ast = \hat{F}_{\textup{SN}}^{-1} (\Phi(U_t^\ast))$;
\item Simulate $\boldsymbol{\nu}_t^\ast \sim \textup{N}(\boldsymbol{0}, \hat{\pmb{\mathsf{\Sigma}}}_t)$;
\item Set $Z_{st}^\ast = \eta_t^\ast  + \nu_{st}^\ast$. 
\end{enumerate}
Here, the split normal variables are simulated using \texttt{qsplitnorm()} from the \texttt{R} package \texttt{fanplot} \citep{Abel2015} and the multivariate normal simulation is carried out by \texttt{mvrnorm()} from the \texttt{R} package \texttt{MASS} \citep{VenablesRipley2002}.

\subsubsection{Adding a climate change signal}\label{subsubsec:downscaling}

We obtain a realization of a stationary climate for the test period corresponding to the mean climate in the training period by setting $X_{st}^\ast = Z_{st}^\ast \hat{\sigma}_{st} + \hat{\mu}_{st}^\ast$, where $Z_{st}^\ast$ is obtained with the simulation algorithm above, $\hat{\sigma}_{st}$ is the standard deviation estimate in \eqref{eq:Z} and $\hat{\mu}_{st}^\ast$ is the mean estimate in \eqref{eq:Z} without the trend component centered such that $\bar{\hat{\mu}}_{s\cdot}^\ast = \bar{\hat{\mu}}_{s\cdot}$. We call this realization \textbf{Xstar} and it serves as a reference for the other methods described below.     

In the realization \textbf{XstarTrend} defined by $X_{st}^{\text{Trend}} = Z_{st}^\ast \hat{\sigma}_{st} + \hat{\mu}_{st}^{\text{Trend}}$ adjustments are made to the baseline and the linear trend components of the mean to reflect the RCM-simulated changes from the test period to the training period. Here, we assume that the long-term average changes at RCM scale directly carry over to the seNorge scale using the information from the RCM gird cell that has the largest intersection area with the seNorge grid cell. We further investigated adding RCM-simulated changes in the seasonality of the mean. However, this resulted in substantially reduced agreement between our model and the out-of-sample data.

Finally, we generate the realization \textbf{XstarTrendVar} with $X_{st}^{\text{TrendVar}} = Z_{st}^\ast \hat{\sigma}^{\text{Var}}_{st} + \hat{\mu}_{st}^{\text{Trend}}$ where adjustment is made to both the mean and the variance. While the mean is adjusted as before, the variance at the seNorge grid level is adjusted such that changes compared to the training period at the upscaled RCM grid level match those in the corrected RCM output.

\subsection{Reference method} \label{subsec:eqm}
We compare the final results from our method to empirical quantile mapping (\textbf{EQM}) \citep[e.g.][]{Piani&2010, Gudmundsson&2012}, a widely adopted method for bias-correcting and downscaling RCM outputs to a finer grid. The EQM method utilizes the empirical cumulative distribution function (eCDF) for variables at both scales. In a first step, we re-grid the RCM output to the seNorge grid using a simple nearest neighbor method. Then, we derive a transfer function matching the RCM-scale eCDF with the seNorge-scale eCDF. The eCDFs are approximated using tables of empirical percentiles with fixed interval of 0.1 spanning the probability space $[0, 1]$. Spline interpolation is performed for the values in between these percentiles and to extrapolate beyond the highest and lowest observed values. In the training period, we derive twelve calendar-month-specific transfer functions for each seNorge grid cell. These transfer functions are assumed to be valid for use in the test period. And we apply them to adjust the RCM output quantile by quantile so that they yield a better match with the seNorge data. To perform the EQM, we employ the \texttt{R} package \texttt{qmap} version 1.0-4 \citep{qmap2016}.

\subsection{Evaluation methods}\label{subsec:evaluation}

We assess the performance of the post-processing methods by comparing projections to out-of-sample data. We compare the marginal distributions in each grid cell using eCDFs over all time points in the test period, the temporal autocorrelation and the spatial correlation in each catchment.  

We compare two marginal distributions $F$ and $G$ using the integrated quadratic distance \citep[IQD;][]{Thorarinsdottir&2013},
\begin{equation}\label{eq:IQD}
  \textup{IQD}(F,G,\omega) = \int_{-\infty}^{+\infty} (F(x) - G(x))^2 \omega(x) \textup{d}x,
\end{equation}
where $\omega$ denotes a non-negative weight function that can be designed to focus on particular part of the distributions. Here, we consider four different weighting options, the unweighted version with $\omega_1 \equiv 1$, as well as weights that focus on the tails and the center of the distributions. Specifically, we set 
\begin{align}
  \omega_2(x) & = \mathbbm{1} \{ x \ge G^{-1}(0.95) \} \label{eq:omega2},\\
  \omega_3(x) & = \mathbbm{1} \{ G^{-1}(0.45) \le x \le G^{-1}(0.55) \}, \label{eq:omega3} \\
  \omega_4(x) & = \mathbbm{1} \{ x \le G^{-1}(0.05) \},  \label{eq:omega4}
\end{align}
where $\mathbbm{1}\{ x \geq u\}$ denotes the indicator function that is equal to one if $x \geq u$ and zero otherwise, and $G$ denotes the data eCDF. Here, a lower IQD value indicates a better correspondence between $F$ and $G$ and we report average IQD values across all grid cells in a catchment,
\begin{equation}
\frac{1}{S} \sum_{s=1}^S \textup{IQD}(F_s, G_s, \omega),
\end{equation}
together with uncertainty bounds obtained by bootstrapping.  

For distributions $F$ and $G$ with finite first moments, the IQD is the score divergence of the continuous ranked probability score (CRPS) which is a proper scoring rule \citep{GneitingRaftery2007}. It thus fulfills a similar propriety condition and can be used to rank competing methods \citep{Thorarinsdottir&2013}. In fact, using the IQD in \eqref{eq:IQD} will result in the same model rankings as computing the average CRPS over all the observations in $G$. However, we find that using the IQD provides improved interpretability as the lowest possible IQD value is zero if $F = G$ while the lowest possible CRPS value depends on the unknown true data distribution. 

At the seNorge scale, the assessment of temporal and spatial correlation structures is carried out separately in each catchment. For the temporal correlation, we aggregate the daily gridded data into a single time series and, subsequently, calculate the autocorrelation up to a certain lag using the function \texttt{Acf()} from the \texttt{R} package \texttt{forecast} \citep{Hyndman&2018}. For the spatial correlation, we calculate the empirical semi-variogram for each month using the same \texttt{R} functions as described in Section~\ref{subsubsec:statisticalModel}.

%%%%%%%%%%%%%%%%
%% Results
%%%%%%%%%%%%%%%%
\section{Results}\label{sec:results}

\subsection{Bias correction at model scale} \label{subsec:resultsBiasCorrection}

\begin{figure}[H]
	\begin{center}
		\includegraphics[width=0.6\textwidth]{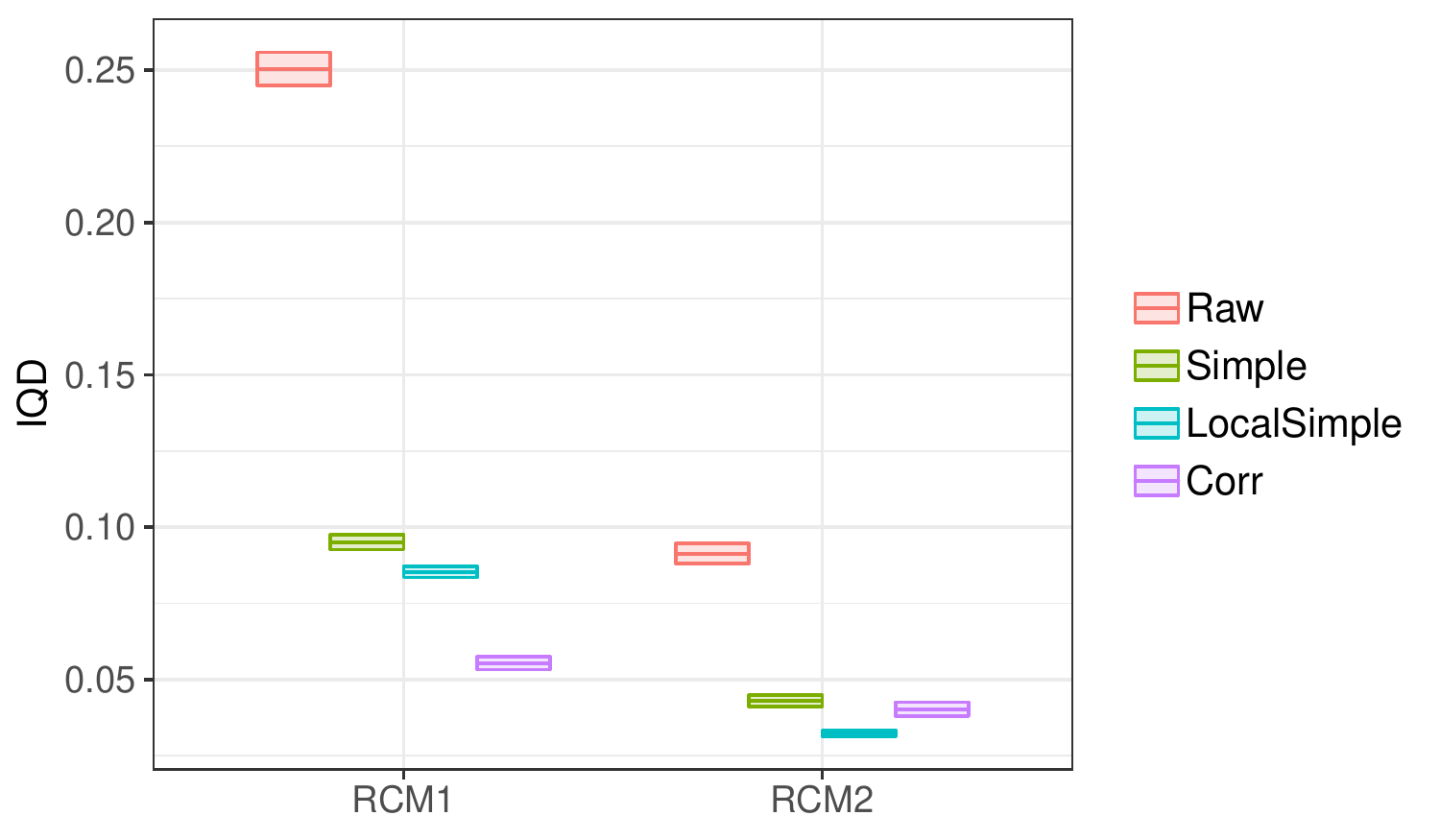}
		\caption{Marginal performance of RCM raw output and three bias correction methods aggregated over the RCM grid cells in the study area, as measured by the integrated quadratic distance (IQD) with a lower value indicating better performance. The marginal distribution over all days in 1987-2005 is compared to the corresponding distribution derived from the upscaled seNorge data product. Raw output and each bias correction method is indicated by color. The middle line of a crossbar indicates the average IQD value across the grid cells while the lower and upper bounds indicate a $90\%$ score uncertainty obtained with 100000 bootstrap samples.}
		\label{fig:IQDRCM}
	\end{center}
\end{figure}

The marginal performance of the three bias correction methods at RCM model scale is shown in Figure~\ref{fig:IQDRCM}. There is notable difference in performance between the two raw RCM outputs with RCM2 more compatible with the upscaled seNorge data. A considerable decrease in the IQD from \textbf{Raw} to \textbf{Simple} means that both RCMs fail to capture the correct long-term average over the whole study area, as expected from free running climate models. Additional improvement can be achieved by local mean correction (\textbf{LocalSimple}). For RCM1, the proposed bias correction method \textbf{Corr} further improves the compatibility with the data product. For RCM2, however, \textbf{Corr} performs worse than \textbf{LocalSimple}, indicating that additional correction of the variance, the linear trend and seasonality of the mean has a slightly adverse effect for RCM2.

\begin{figure}[H]
  \begin{center}
    \includegraphics[width=0.85\textwidth]{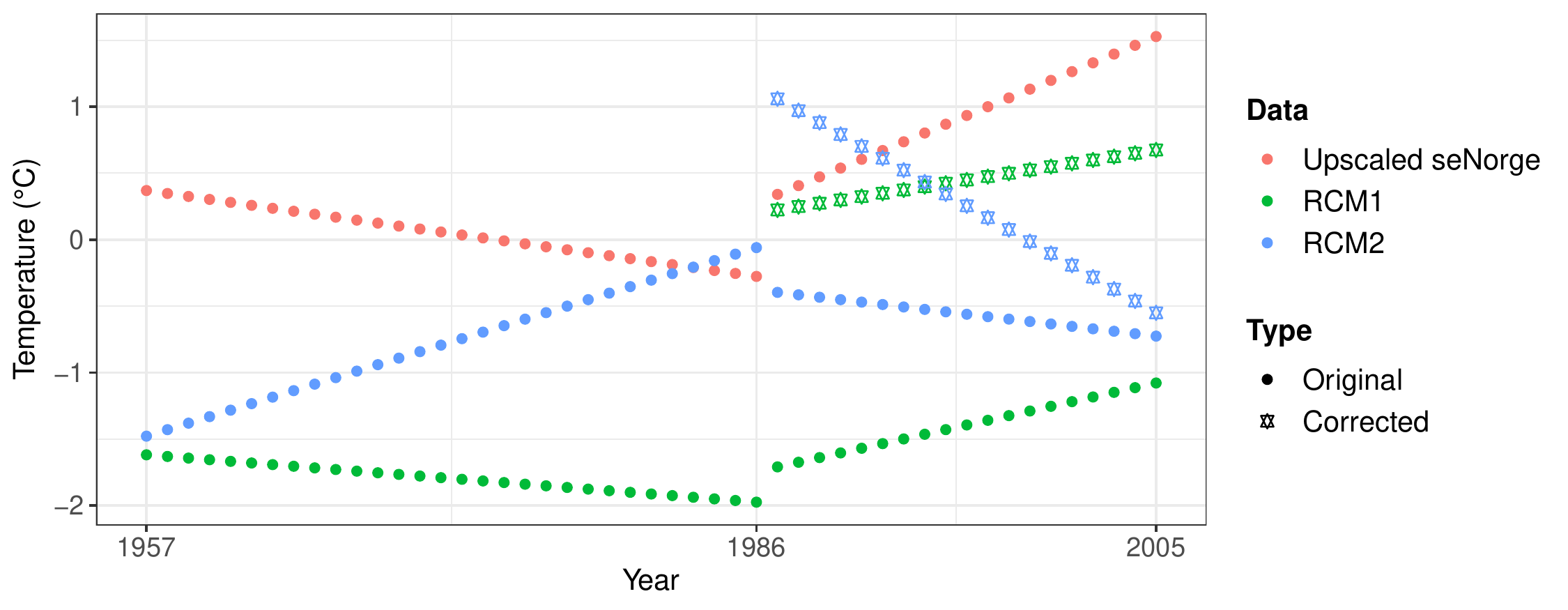}
    \caption{Combined baseline (\ref{eq:f1}) and linear trend (\ref{eq:f3}) components of the estimated mean for one RCM grid cell in the study area, for the upscaled seNorge data product and the two RCMs over the training period 1957-1986 (left) and the test period 1987-2005 (right), where also the corrected estimates of the two RCMs are indicated. The estimates are standardized such that the overall mean of the data product in the training period equals 0.}
    \label{fig:trend}
  \end{center}
\end{figure}

To investigate this further, consider the estimated mean baseline and trend for a single grid cell shown in Figure~\ref{fig:trend}. The upscaled data product has a slightly negative trend in the training period 1957-1986 and a positive trend in the test period 1987-2005 with the overall mean temperature in the test period 0.9$^\circ$C higher than that in the training period. While RCM1 has a baseline estimate that is around 2$^\circ$C colder than the data product, the trend estimates of the two data sets are similar, resulting in a bias-corrected RCM output that is overall 0.55$^\circ$C colder than the data product with a slightly slower warming rate. The raw output from RCM2, on the other hand, has opposite trends compared to the data product in both time periods, resulting in a bias-corrected trend that further exaggerates the model errors. As a result, the empirical distribution function over all time points in the test period for the \textbf{Corr} method has a much larger spread than that for the data product or that obtained under \textbf{LocalSimple}.  

\subsection{Bias correction and downscaling} \label{subsec:resultsSD}

\subsubsection{Marginal performance} \label{subsubsec:resultsSDIQD}

\begin{figure}[H]
	\begin{center}
		\includegraphics[width=0.85\textwidth]{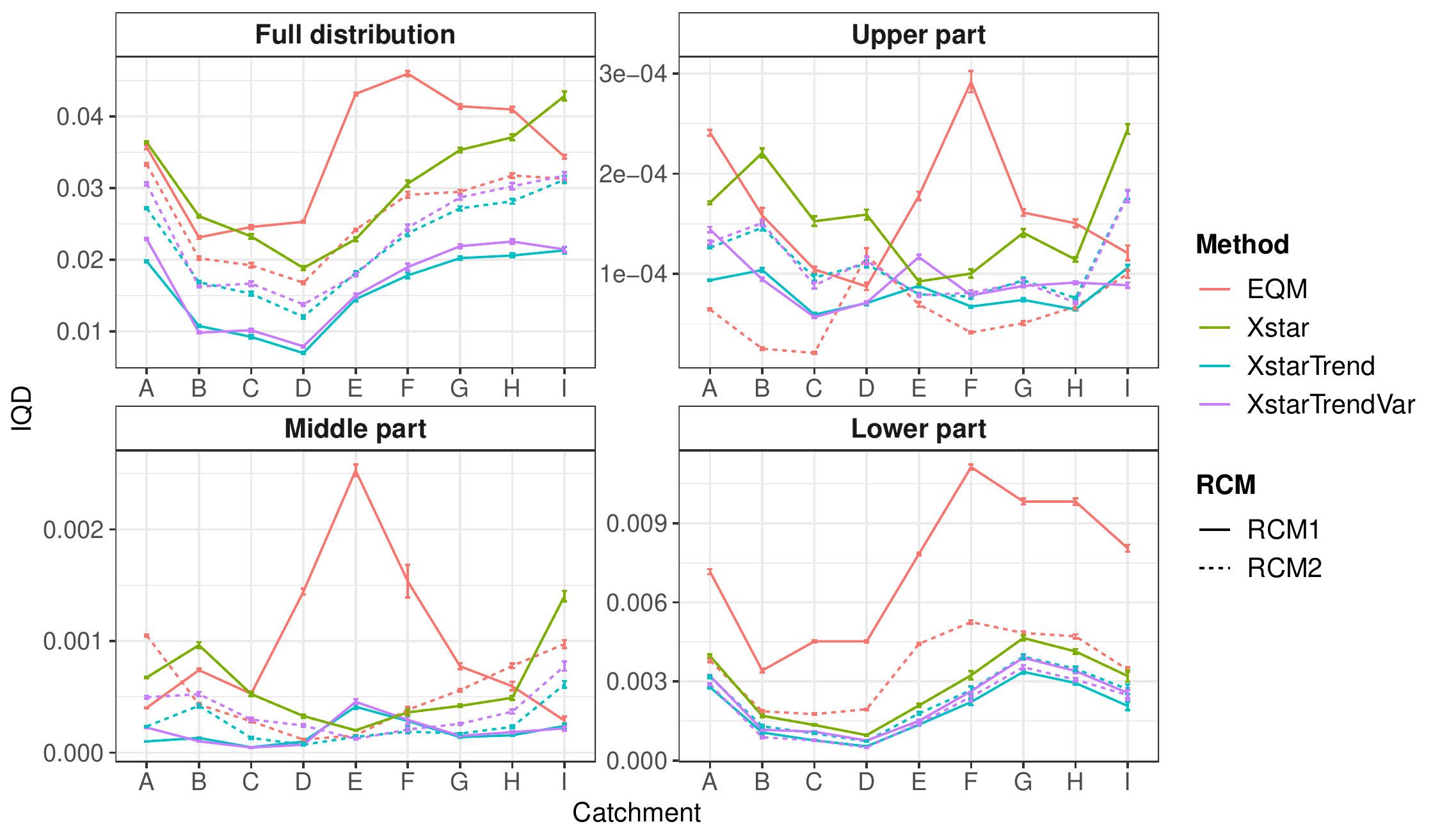}
		\caption{Integrated quadratic distance (IQD) values for marginal comparison of the daily seNorge data product and post-processed RCM model output 1987-2005 aggregated over the grid cells in each catchment. A lower value indicates a better performance. Post-processing method is indicated by color and RCM by line type. The full distributions are compared in the top left plot while comparisons focusing on the upper part (\ref{eq:omega2}), middle part (\ref{eq:omega3}) and lower part (\ref{eq:omega4}) of the distributions are also shown. The score uncertainty is indicated with a $90\%$ uncertainty bound obtained with 100000 bootstrap samples.}
	\label{fig:IQDcat}
	\end{center}
\end{figure}

The marginal performance at the fine scale is assessed in Figure~\ref{fig:IQDcat}. From the example in Figure~\ref{fig:trend}, we see that the mean climate varies between the two time periods. The results here similarly show that adding climate change information from the RCMs substantially improves the WG realization centered on the mean climate in the training period (denoted \textbf{Xstar}). For each RCM, the difference between \textbf{XstarTrend} and \textbf{XstarTrendVar} is small, the mean-only correction of \textbf{XstarTrend} commonly showing minimally better performance. This may partly be explained by the fact that there is little difference between the variances of the two time periods, and partly by the fact that the coarse resolution variance of the RCMs does not perfectly relate to the fine scale variance of the data product. We obtain consistently better results using the climate change information from RCM1 than RCM2. This is in line with the trend estimation results shown in Figure~\ref{fig:trend}, despite the bias-corrected RCM2 showing better overall marginal performance at the model scale, cf. Figure~\ref{fig:IQDRCM}. 

The proposed two-step post-processing approach shows consistently better marginal performance than EQM under an assessment of the full distribution and when focusing on the lower tail. When focusing on the upper tail, or the central part of the distribution, the differences between the methods are smaller and the method ranking varies substantially across the catchments. Notably, the EQM results are much better for RCM2 than for RCM1. Furthermore, the EQM performance is less stable across the different catchments with some indications of worse performance in the inland catchments A and E through I. 

As mentioned in Section \ref{subsubsec:statisticalModel}, we have also investigated a slightly simpler two-step post-processing procedure where the temporal residual series is assumed to follow an $\textup{ARMA}(p, q)$ model with Gaussian marginals. This simplification results in significantly reduced performance, adding approximately 0.008 to the average IQD value of the full distribution per catchment (results not shown). For RCM2, EQM performs better than this simplified two-step approach in seven out of the nine catchments. 

Below, we further assess the spatial and temporal characteristics of the \textbf{XstarTrend} method applied to RCM1 and EQM applied to RCM2.

\subsubsection{Spatio-temporal dependence structure} \label{subsubsec: STdependence}

\begin{figure}[H]
	\begin{center}
		\includegraphics[width=0.75\textwidth]{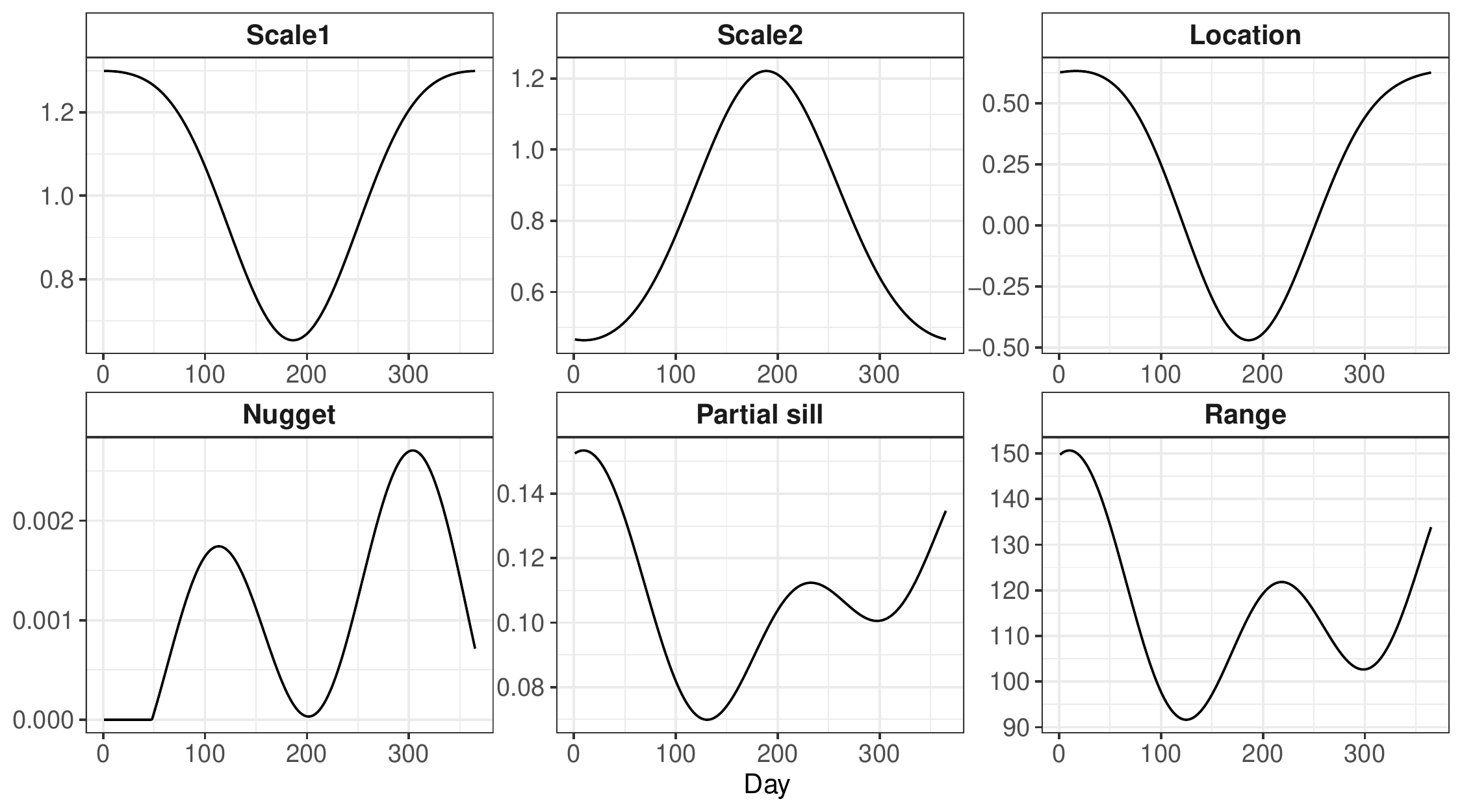}
		\caption{Parameter estimates in catchment A, Gaulfoss, in the training period 1957-1986 for the residual models in \eqref{eq:SN} and \eqref{eq:Cov}. Top row: The two scale parameters, $\sigma_{1t}, \sigma_{2t}$, and the location parameter $\mu_t$ of the split normal distribution in equation (\ref{eq:SN}). Bottom row: The parameters of the exponential covariance function in (\ref{eq:Cov}), nugget $\theta_{0t}$, partial sill $\theta_{1t}$ and range $\theta_{2t}$.}
		\label{fig:parGaulfoss}
	\end{center}
\end{figure}

Parameter estimates for the split normal residual model in \eqref{eq:SN} and the spatial covariance function in \eqref{eq:Cov} in the largest catchment A, Gaulfoss, are given in Figure~\ref{fig:parGaulfoss}. The parameters of the split normal distribution follow a seasonal pattern that can be deducted from the data plot in Figure~\ref{fig:boxplotEta}, with the scale parameter for the lower tail, $\sigma_{1t}$, being higher in winter and lower in summer and the opposite holding for $\sigma_{2t}$, the scale parameter for the upper tail. While the location parameter estimates are, by construction, approximately mean zero over the entire year, these also follow a seasonal pattern with negative values in summer and positive values in winter. The spatial covariance function similarly exhibits a seasonal pattern. The spatial correlation has the highest range in winter followed by summer, while the range is smaller in spring and fall when, instead, the nugget parameter takes positive values. 

\begin{figure}[H]
	\begin{center}
		\includegraphics[width=0.75\textwidth]{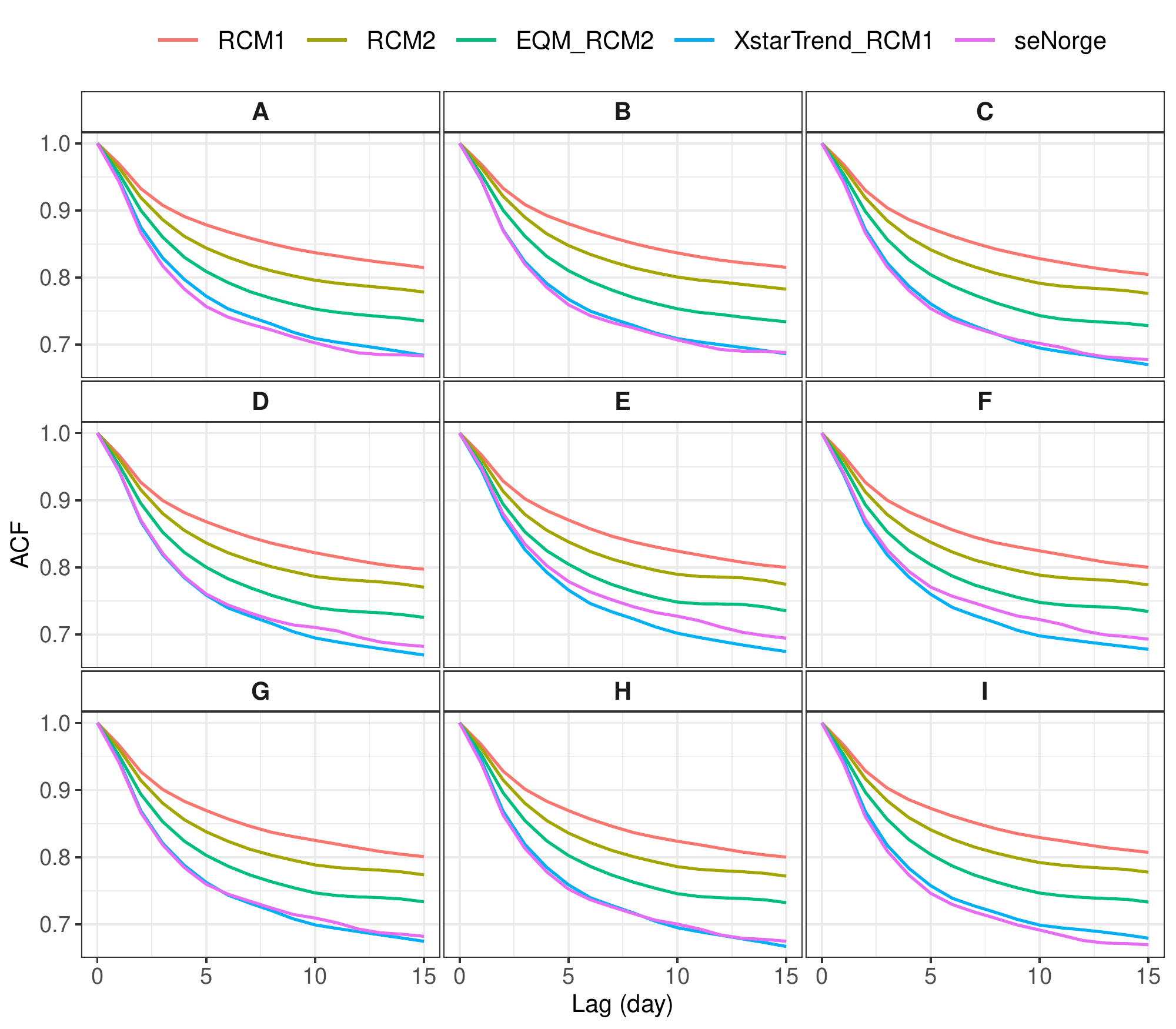} 
		\caption{The temporal dependence in the nine catchments, measured by an autocorrelation function (ACF) of the average time series over the daily fields from 1987-2005, for the raw RCM output and two downscaling methods is compared to that from the seNorge data product.}
		\label{fig:ACF}
	\end{center}
\end{figure}

To assess the temporal dependence, we estimate the autocorrelation function of the average daily temperature series in each catchment, see Figure~\ref{fig:ACF}. The results are very similar across the catchments: The raw RCM output has a substantially higher autocorrelation than the finer-scale seNorge data product and even if this is somewhat corrected in EQM, the results are not quite comparable to seNorge. The \textbf{XstarTrend} post-processing inherits its temporal dependence structure mostly from the seNorge data product in training period, resulting in temporal dependence very similar to that of the data product in the test period.  

\begin{figure}[H]
	\begin{center}
		\includegraphics[width=0.8\textwidth]{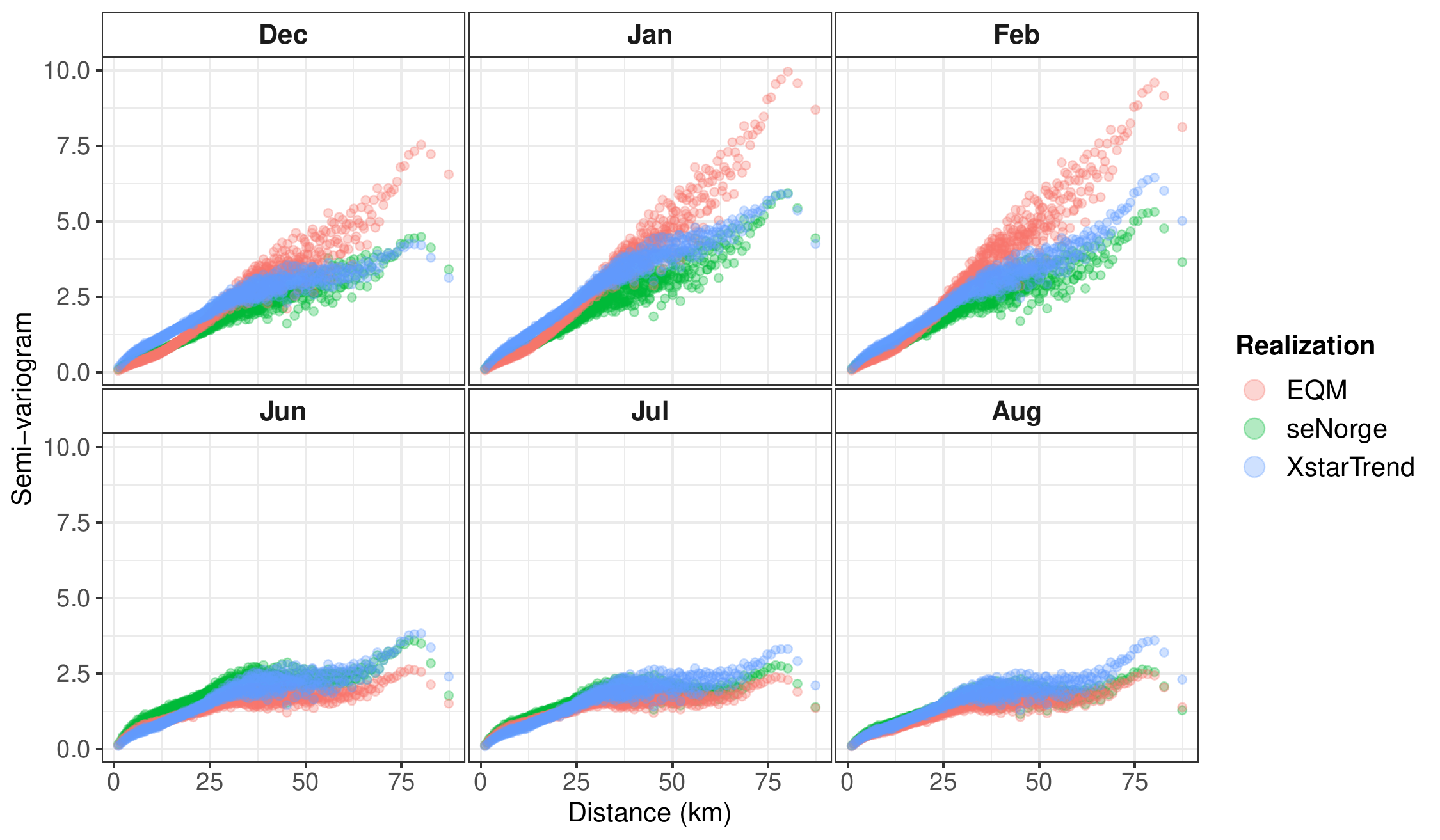}
		\caption{Spatial dependence in catchment A (Gaulfoss) in 1987-2005 by month, as measured by an empirical semi-variogram based on the daily temperature fields. The plots show empirical semi-variograms derived from the seNorge data product and two downscaled results, EQM applied to RCM2 and XstarTrend applied to RCM1.}
		\label{fig:spaceCorGaulfoss}
	\end{center}
\end{figure}

For assessing the spatial dependence structure, we focus on the largest catchment A, Gaulfoss. Figure \ref{fig:spaceCorGaulfoss} shows the empirical semi-variograms for the seNorge data product, the \textbf{XstarTrend} method applied to RCM1 and EQM applied to RCM2 in the winter and summer months. While all three methods are comparable in the summer, the spatial dependence in winter is better modeled by \textbf{XstarTrend} than EQM. Due to its continental climate, the spatial dependence in the temperature at Gaulfoss is quite different for the two seasons. The $y$-value attained when the semi-variogram starts to level off, called {\em sill} in geostatistics, measures the total variance of the variable within the spatial domain. The temperatures are more variable in winter, leading to a larger sill, cf. Figure~\ref{fig:parGaulfoss}. This feature is properly captured by \textbf{XstarTrend}, and largely overestimated by EQM.  The distance where the semi-variogram first flattens out, the {\em range}, is typically around 30-35 km in summer and somewhat longer in winter, potentially due to dominance of continental arctic air masses in the region. In winter, the semi-variogram values given by \textbf{XstarTrend} level off similarly as the seNorge data, whereas those by EQM show substantial differences for distances longer than 30 km. Note that these patterns are somewhat different for the other eight catchments (results not shown). 

\begin{figure}[H]
	\begin{center}
		\includegraphics[width=0.9\textwidth]{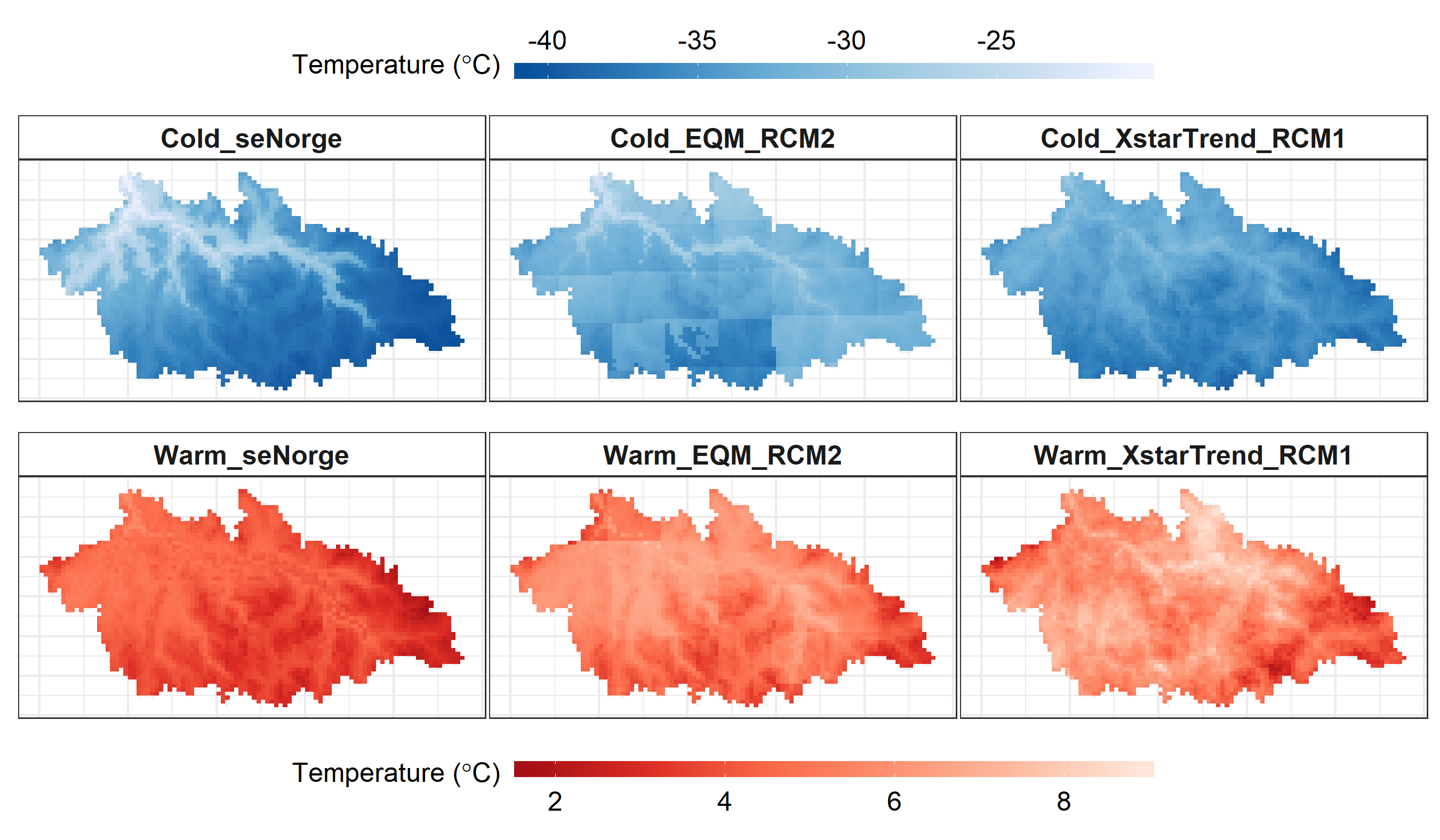} 
		\caption{Examples of the coldest (top row) and warmest (bottom row) temperature fields in catchment A (Gaulfoss) over all January days in the test period 1987-2005. The examples shown are from the seNorge data product, EQM applied to RCM2 and XstarTrend applied to RCM1.}
		\label{fig:coldWarmJan}
	\end{center}
\end{figure}

Figure \ref{fig:coldWarmJan} shows examples of cold and warm January days from the seNorge data product and the two post-processing methods. EQM is applied independently for each RCM grid cell and it can be seen that the resulting daily temperature fields have artificial boundaries corresponding to the RCM grid cells, while those by \textbf{XstarTrend} do not have such boundaries and show a spatial consistency closer to the seNorge temperature fields.  

%%%%%%%%%%%%%%%%
%% Discussion and conclusions
%%%%%%%%%%%%%%%%
\section{Discussion and Conclusions}\label{sec:conclusions}

We propose a two-step statistical post-processing procedure that bias-corrects and downscales RCM simulations to a high-resolution grid. Our objective is to develop a full-field downscaling method for daily mean temperature that explicitly accounts for the fine-scale variability and dependence in both space and time. Employing two RCMs from the EURO-CORDEX ensemble and the high-resolution gridded observational data product seNorge, we apply the procedure in the Tr{\o}ndelag area of Norway, and find that the generated results are closer to the gridded reference data in terms of marginal, temporal and spatial properties than an empirical quantile mapping (EQM) approach.

Our specific implementation separates statistical bias correction and stochastic downscaling. In a first step, to overcome the representativeness issue of the RCM simulations \citep{MaraunWidmann2018} for local-scale climate, we follow \citet{Volosciuk&2017} and perform bias correction only at the model scale. Then we follow e.g. \citet{Piani&2010} and assume a Gaussian distribution for daily mean temperature. Here, using a model which parameters vary smoothly in space and time, we are able to account for the spatial and day-to-day variation of the two moments as well as a potential linear trend. Calibration is performed once for the full training data set in each catchment. Other post-processing methods, however, often calibrate a model at single locations \citep[e.g.][]{Volosciuk&2017} for individual months (e.g. the EQM applied in current study) or seasons \citep[e.g.][]{Vrac&2012, Wong&2014}. Such separation of the data in space and time may overlook systematic variations with topography or seasons.  Furthermore, they are typically unable to estimate a single long-term linear trend for the whole domain, e.g. separating it from the seasonal variations, and, as a consequence, modify the trend when correcting other properties \citep{MaraunWidmann2018}. 

In a second step, we model the space-time variability at the finer scale using residuals generated on a high-resolution grid for limited areas (hydrological catchments in our case). Even for daily mean temperature which can be assumed Gaussian, simultaneous modeling of the space-time dependence is not straightforward. For computational feasibility and flexibility, we assume stationarity and space-time separability in the model. For the spatial dependence, we employ a similar approach as \citet{Wilks2009}, and specify a parametric covariance function of the exponential type with parameters smoothly changing to describe spatial structure variations through out the year. Alternative, more advanced approaches here include the Mat\'ern covariance model \citep[e.g.][]{Lindgren&2011} or models based on non-parametric approaches such as principal component analysis \citep{Heinrich&2019}. For the temporal dependence, we found that the daily mean residuals have negative skewness in winter and positive skewness in summer \citep[found also in e.g.][]{Huybers&2014}, which if not accounted for would lead to reduced performance of the downscaling methods. Our solution is to combine a split normal distribution for the asymmetry with an ARMA model for the temporal correlation, a computationally feasible approach to non-Gaussian modeling of the temporal process. 

The gap between the bias correction and the stochastic modeling is bridged by adding climate change signal derived from the model scale to the fine-scale WG realizations. Climate change signals in the mean and the variance can be selectively added to form the final results of the proposed procedure. Here, we compared three options: Using just the stationary climate (\textbf{Xstar}), adjusting only the mean (\textbf{XstarTrend}) and adjusting both mean and variance (\textbf{XstarTrendVar}). To assess the agreement between the generated results and the gridded data in the test period, we employ the integrated quadratic distance (IQD) for evaluation of the marginal aspect, the autocorrelation function (ACF) for temporal dependence and the empirical semi-variogram for spatial dependence. We find that in all the catchments in our study area and under both RCMs, \textbf{XstarTrend} and \textbf{XstarTrendVar} perform better than EQM in terms of marginal distribution and temporal dependence, while properly representing spatial dependence. In addition, we found that the skill of an RCM at the coarser scale may not necessarily carry over to the finer scale and agree with \citet{MaraunWidmann2018} that it is important to assess the skill of the climate model output in terms of the information to be used.

%%%%%%%%%%%%%%%%%%%%%%
% ACKNOWLEDGMENTS
%%%%%%%%%%%%%%%%%%%%%%
\section*{Acknowledgments}
This work was supported by the Research Council of Norway through project nr. 255517 ``Post-processing Climate Projection Output for Key Users in Norway''.

%%%%%%%%%%%%%%%%%%%%%%
% REFERENCES
%%%%%%%%%%%%%%%%%%%%%%
\bibliographystyle{ametsoc2014}
\bibliography{references}

\end{document}